\documentclass[12pt,letterpaper]{article}
\usepackage{amsmath,amsfonts,amsthm,amssymb,stmaryrd,bm,cite} 
\usepackage{relsize,tabls,tikz} 
\usepackage{graphicx} 
\usepackage{array}   

\allowdisplaybreaks  

\theoremstyle{plain}

\numberwithin{equation}{section}
\newtheorem{thm}{Theorem}[section]
\newtheorem{lem}[thm]{Lemma}
\newtheorem{cor}[thm]{Corollary}

\newcommand{\integers}{{\mathbb Z}}

\newcommand{\complex}{{\mathbb C}}
\newcommand{\cscript}{{\mathcal C}}
\newcommand{\hscript}{{\mathcal H}}
\newcommand{\sscript}{{\mathcal S}}
\newcommand{\rmprob}{\mathop{Prob}}
\newcommand{\sscripthat}{\widehat{\sscript}}
\newcommand{\overv}{\overline{v}}
\newcommand{\bfe}{\mathbf{e}}    
\newcommand{\bff}{\mathbf{f}}       
\newcommand{\bfg}{\mathbf{g}}       
\newcommand{\bfk}{\mathbf{k}}
\newcommand{\bfp}{\mathbf{p}}       
\newcommand{\bfq}{\mathbf{q}}
\newcommand{\bfx}{\mathbf{x}}       
\newcommand{\bfy}{\mathbf{y}}       
\newcommand{\bfzero}{\mathbf{0}}    
\newcommand{\ctimes}{\mathrel{\mathlarger\cdot}}
  
\newcommand{\ab}[1]{\left|#1\right|}
\newcommand{\doubleab}[1]{\left|\left|#1\right|\right|}
\newcommand{\brac}[1]{\left\{#1\right\}}
\newcommand{\paren}[1]{\left(#1\right)}
\newcommand{\sqbrac}[1]{\left[#1\right]}
\newcommand{\elbows}[1]{{\left\langle#1\right\rangle}}
\newcommand{\ket}[1]{{\left|#1\right>}}
\newcommand{\bra}[1]{{\left<#1\right|}}

\begin{document}
\title{DISCRETE SCALAR\\QUANTUM FIELD THEORY
}
\author{S. Gudder\\ Department of Mathematics\\
University of Denver\\ Denver, Colorado 80208, U.S.A.\\
sgudder@du.edu
}
\date{}
\maketitle

\begin{abstract}
We begin with a description of spacetime by a 4-dimensional cubic lattice $\sscript$. It follows from this framework that the the speed of light is the only nonzero instantaneous speed for a particle. The dual space $\sscripthat$ corresponds to a cubic lattice of energy-momentum. This description implies that there is a discrete set of possible particle masses. We then define discrete scalar quantum fields on $\sscript$. These fields are employed to define interaction Hamiltonians and scattering operators. Although the scattering operator $S$ cannot be computed exactly, approximations are possible. Whether $S$ is unitary is an unsolved problem. Besides the definitions of these operators, our main assumption is conservation of energy-momentum for a scattering process. This article concludes with various examples of perturbation approximations. These include simplified versions of electron-electron and electron-proton scattering as well as simple decay processes. We also define scattering cross-sections, decay rates and lifetimes within this formalism.
\end{abstract}\newpage

\section{Introduction}  
It has long been recognized that standard quantum field theory suffers from a plague of singularities and infinities. The author would even go so far as to say that in its usual form, a quantum field does not exist! By this we mean that it does not have a mathematically well-defined definition. The problems are compounded by employing these nonexistent quantum fields in rather complicated ways to define interaction Hamiltonians and scattering operators. Perturbation methods are then used to compute numbers that frequently agree with experiment. However, to obtain these numbers, renormalization techniques and infinity cancellations have to be employed
\cite{ps95,vel94}. Attempts have been made to develop a rigorous mathematically well-define quantum field theory \cite{sw64}. These attempts have been successful for free quantum fields, but they have not adequately included interacting fields with nontrivial scattering.

It seems to us that the best way to overcome these difficulties is to postulate that spacetime is discrete \cite{gud68,hei30,rus54}. Although such a granular structure for spacetime has not been experimentally observed, with increasingly accurate instruments it may become evident in the future, perhaps in an indirect manner. The granular structure may manifest itself in terms of elementary lengths and times at Planck scales of about $10^{-33}$cm. and $10^{-43}$sec., respectively. The simplest such framework would be a rigid cubic 4-dimensional lattice. The author has previously studied a tetrahedral lattice that appears to have certain advantages \cite{gud16} but for simplicity and ease of computation, we shall only consider the cubic lattice $\sscript$ here.

It follows from this description that the speed of light is the only nonzero instantaneous speed for a particle. We observe slower speeds because we actually measure average speeds of objects. The dual space $\sscripthat$ corresponds to a 4-dimensional cubic lattice of energy-momentum. This description implies that there is a discrete set of possible particle masses, which we then determine.

We next define discrete scalar quantum fields on $\sscript$. We show that these fields exist mathematically and derive some of their properties. We mention one can also study discrete vector quantum fields that involve spin, but  for simplicity we only consider the spin zero case here. Various fields are combined to form interaction Hamiltonians. We postulate that the scattering operator satisfies a ``second quantization'' Schr\"odinger equation and derive its form. Besides the definition of these operators, our main assumption is conservation of energy-momentum for scattering processes. Due to this conservation law, we show that a scattering process usually possesses only a finite number of possible outgoing states. These can then be summed over to find scattering amplitudes and probabilities.

The article concludes with various examples of perturbation approximations. These include simplified versions of electron-electron and electron-proton scattering as well as simple decay processes. We also define scattering cross-sections, decay rates and lifetimes within this formalism. We compute these quantities in simple cases. Finally, some speculations are made about the theory's ability to predict the existence of dark energy and dark matter.

\section{Discrete Spacetime and Energy-Momentum} 
Our basic assumption is that spacetime is discrete and has the structure of a 4-dimensional cubic lattice $\sscript$. We regard $\sscript$ as a framework or scaffolding in which the vertices of the lattice (or network) represent tiny cells of Planck scale that may or may not be occupied by a particle. The edges between vertices represent space directions in which particles can propagate at a given time. Let
\begin{equation*}
\integers =\brac{0,\pm 1,\pm 2,\ldots}
\end{equation*}
be the integers and $\integers ^+=\brac{0,1,2,\ldots}$ be the nonnegative integers. We have that $\sscript =\integers ^+\times\integers ^3$ where $\integers ^+$ represents discrete time and $\integers ^3$ represents discrete 3-space. If $x=(x_0,x_1,x_2,x_3)\in\sscript$, we write $x=(x_0,\bfx )$ where $x_0\in\integers ^+$ is time and
$\bfx\in\integers ^3$ is a 3-space point. We have that $\sscript$ is a module in the sense that $\sscript$ is closed under addition and multiplication by elements of
$\integers ^+$. The vectors
\begin{align*}
d&=(1,\bfzero )=(1,0,0,0),\enspace e=(0,\bfe )=(0,1,0,0)\\
f&=(0,\bff )=(0,0,1,0),\enspace g=(0,\bfg )=(0,0,0,1)
\end{align*}
form a basis for $\sscript$ and every $x\in\sscript$ has the unique form
\begin{equation*}
x=nd+me+pf+qg
\end{equation*}
$n\in\integers ^+$, $m,p,q\in\integers$.

We equip $\sscript$ with the Minkowski distance
\begin{equation*}
\doubleab{x}_4^2=x_0^2-\doubleab{\bfx}_3^2=x_0^2-x_1^2-x_2^2-x_3^2
\end{equation*}
As usual $x,y\in\sscript$ are \textit{time-like separated} if $\doubleab{x-y}_4^2\ge 0$ and \textit{space-like separated} if $\doubleab{x-y}_4<0$. For $x\in\sscript$, the set
\begin{equation*}
\cscript ^+(x)=\brac{y\in\sscript\colon y_0\ge x_0,\doubleab{y-x}_4^2\ge 0}
\end{equation*}
is the \textit{future light cone at} $x$. Of course, we are using units in which the speed of light is 1. With our interpretation of the structure of $\sscript$, it seems natural that a particle at $\bfx$ can either stay at $\bfx$ or move in one of the six space directions to a point $\bfy$ in one time unit, where
$\bfy =\bfx\pm\bfe, \bfx\pm\bff ,\bfx\pm\bfg$. In this way, the only nonzero instantaneous speed of a particle is the speed of light~1. This gives a primitive reason why the speed of light is the upper limit for all signal speeds. The reason that we observe slower speeds is that we usually measure average speeds and not instantaneous ones. If a particle moves from $\bfx$ to $\bfy$, its \textit{average speed} is $\overv =\doubleab{\bfy -\bfx}_3/t$ where $t$ is the elapsed proper time; that is, the time measured along the particle trajectory. For example, suppose a particle initially at $\bfx$, stays at $\bfx$ for one time unit and then moves to $\bfy =\bfx +\bfe$ in the next time unit. Then the proper time $t=2$ and the average speed is $\overv=\doubleab{\bfe}_3/2=1/2$. As another example, suppose a particle initially at $\bfzero$ moves to $\bfe$ at the first time unit and then to $\bfe +\bff$ at the second time unit. The proper time is again $t=2$ and $\overv=\doubleab{\bfe +\bff}_3/2=1/\sqrt{2}$. We conclude that photons must travel in straight lines along coordinate axes.

The dual of $\sscript$ is denoted by $\sscripthat$. We regard $\sscripthat$ as having the identical structure as $\sscript$ and that $\sscripthat$ again has basis $d,e,f,g$.
The only difference is that we denote elements of $\sscripthat$ by
\begin{equation*}
p=(p_0,\bfp )=(p_0,p_1,p_2,p_3)
\end{equation*}
and interpret $p$ as the energy-momentum vector for a particle. In fact, we sometimes even call $p\in\sscripthat$ a particle. Moreover, we only consider the forward light cone
\begin{equation*}
\cscript ^+(0)=\brac{p\in\sscripthat\colon\doubleab{p}_4\ge 0}
\end{equation*}
in $\sscripthat$. For a particle $p\in\sscripthat$, we call $p_0\ge 0$ the \textit{total energy}, $\doubleab{\bfp}_3\ge 0$ the \textit{kinetic energy}
and $m=\doubleab{p}_4\ge 0$ the \textit{mass} of $p$. The integers $p_1,p_2,p_3$ are \textit{momentum components}. Since
\begin{equation}         
\label{eq21}
m^2=\doubleab{p}_4^2=p_0^2-\doubleab{\bfp}_3^2
\end{equation}
we conclude that Einstein's energy formula $p_0=\sqrt{m^2+\doubleab{\bfp}_3^2}$ holds.

It is clear that any nonnegative integer can be a total energy. However, the square of a kinetic energy must be the sum of three squares
$\doubleab{\bfp}_3^2=p_1^2+p_2^2+p_3^2$. A number theory theorem \cite{vde87} says that $n\in\integers ^+$ is a sum of three squares if and only if $n$ does not have the form $4^k(8s+7)$, $k,s\in\integers ^+$. We conclude that $7,15,23,28,\ldots$ are eliminated. Thus, $\doubleab{\bfp}_3=\sqrt{n}$, $n\in\integers ^+$ is a kinetic energy if and only if $n\ne 7,15,23,28,\ldots\,$. It follows that every nonnegative number $\sqrt{n}$, $n\in\integers ^+$ is a possible mass and vice versa, although it may not appear for certain total energy $p_0\ge\sqrt{n}$. Therefore, this theory predicts that there are a countable number of admissible particle masses. In contrast to the speeds considered previously, we also have the concept of a \textit{geometric speed} for $p\in\sscripthat$ given by
\begin{equation*}
v=\frac{\doubleab{\bfp}_3}{p_0}=\frac{\sqrt{p_0^2-m^2}}{p_0}=\sqrt{1-\frac{m^2}{p_0^2}}
\end{equation*}
whenever $p_0\ne 0$. Of course, $0\le v\le 1$ and $v=1$ if and only if $m=0$. Moreover, $v=0$ if and only if $\doubleab{\bfp}=0$.

Table~1 lists kinetic energy values, the corresponding momentum vectors and the number of such vectors. Notice, as previously observed, that $\sqrt{7}$, $\sqrt{15}$, $\sqrt{23}$ are missing.

We can apply Table~1 to find the possible particle masses $m$ corresponding to total energy $p_0$ as well as the number of particle kinetic energies. In this work, we assume that $p_0\ne 0$.
\newpage

\vglue -5pc
{\parindent=-3pc
\begin{tabular}{c|c|c}
Kinentic&&\hfill\hfill\\
Energy&Momentum Vectors&Number\hfill\hfill\\
\hline
1&$\pm\bfe ,\ \pm\bff ,\ \pm\bfg$\hfill\hfill&6\\\hline
$\sqrt{2}$&$\pm\bfe\pm\bff ,\ \pm\bfe\pm\bfg ,\ \pm\bff\pm\bfg$\hfill\hfill&12\\\hline
$\sqrt{3}$&$\pm\bfe\pm\bff\pm\bfg$\hfill\hfill&8\\\hline
2&$\pm 2\bfe ,\ \pm 2\bff ,\ \pm 2\bfg$\hfill\hfill&6\\\hline
$\sqrt{5}$&$\pm 2\bfe\pm\bff ,\ \pm 2\bfe\pm\bfg ,\ \pm 2\bff\pm\bfe ,\ \pm 2\bff\pm\bfg ,\ \pm 2\bfg\pm\bfe ,\ \pm 2\bfg\pm\bff$\hfill\hfill&24\\\hline
$\sqrt{6}$&$\pm 2\bfe\pm\bff\pm\bfg ,\ \pm2\bff\pm\bfe\pm\bfg ,\ \pm 2\bfg\pm\bfe\pm\bff$\hfill\hfill&24\\\hline
$\sqrt{8}$&$\pm 2\bfe\pm 2\bff ,\ \pm 2\bfe\pm 2\bfg ,\ \pm 2\bff\pm 2\bfg$\hfill\hfill&12\\\hline
3&$\pm 3\bfe ,\ \pm 3\bff ,\ \pm 3\bfg ,\ \pm 2\bfe\pm 2\bff\pm\bfg ,\ \pm 2\bfe\pm 2\bfg\pm\bff ,\ \pm 2\bff\pm 2\bfg\pm\bfe$\hfill\hfill&30\\\hline
$\sqrt{10}$&$\pm 3\bfe\pm\bff ,\ \pm 3\bfe\pm\bfg ,\ \pm 3\bff\pm\bfe ,\ \pm 3\bff\pm\bfg ,\ \pm 3\bfg\pm\bfe ,\ \pm 3\bfg\pm\bff$\hfill\hfill&24\\\hline
$\sqrt{11}$&$\pm 3\bfe\pm\bff\pm\bfg ,\ \pm 3\bff\pm\bfe\pm\bfg ,\ \pm 3\bfg\pm\bfe\pm\bff$\hfill\hfill&24\\\hline
$\sqrt{12}$&$\pm 2\bfe\pm 2\bff\pm 2\bfg$\hfill\hfill&8\\\hline
$\sqrt{13}$&$\pm 3\bfe\pm 2\bff ,\ \pm 3\bfe\pm 2\bfg ,\ \pm 3\bff\pm 2\bfe ,\ \pm 3\bff\pm 2\bfg ,\ \pm 3\bfg\pm 2\bfe ,\ \pm 3\bfg\pm 2\bff$\hfill\hfill&24\\\hline
$\sqrt{14}$&$\pm 3\bfe\pm 2\bff\pm\bfg ,\ \pm 3\bfe\pm 2\bfg\pm\bff ,\ \pm 3\bff\pm 2\bfe\pm\bfg ,\ \pm 3\bff\pm 2\bfg\pm\bfe$,\hfill\hfill&\\
&$\pm 3\bfg\pm 2\bfe\pm\bff ,\ \pm 3\bfg\pm 2\bff\pm\bfe$\hfill\hfill&48\\\hline
4&$\pm 4\bfe ,\ \pm 4\bff ,\ \pm 4\bfg$\hfill\hfill&6\\\hline
$\sqrt{17}$&$\pm 4\bfe\pm\bff ,\ \pm 4\bfe\pm\bfg ,\ \pm 4\bff\pm\bfe ,\ \pm 4\bff\pm\bfg ,\ \pm 4\bfg\pm\bfe ,\ \pm 4\bfg\pm\bff$,\hfill\hfill&\\
&$\pm 3\bfe\pm 2\bff\pm 2\bfg ,\ \pm 3\bff\pm 2\bfe\pm 2\bfg ,\ \pm 3\bfg\pm 2\bfe\pm 2\bff$\hfill\hfill&48\\\hline
$\sqrt{18}$&$\pm 4\bfe\pm\bff\pm\bfg ,\ \pm 4\bff\pm\bfe\pm\bfg ,\ \pm 4\bfg\pm\bfe\pm\bff ,\ \pm 3\bfe\pm 3\bff ,\ \pm 3\bfe\pm 3\bfg ,\ \pm 3\bff\pm 3\bfg$\hfill\hfill&36\\\hline
$\sqrt{19}$&$\pm 3\bfe\pm 3\bff\pm\bfg ,\ \pm 3\bfe\pm 3\bfg\pm\bff ,\ \pm 3\bff\pm 3\bfg\pm\bfe$\hfill\hfill&24\\\hline
$\sqrt{20}$&$\pm 4\bfe\pm 2\bff ,\ \pm 4\bfe\pm 2\bfg ,\ \pm 4\bff\pm 2\bfe ,\pm 4\bff\pm 2\bfg ,\ \pm 4\bfg\pm 2\bfe ,\ \pm 4\bfg\pm 2\bff$\hfill\hfill&24\\\hline
$\sqrt{21}$&$\pm 4\bfe\pm 2\bff\pm\bfg ,\ \pm 4\bfe\pm 2\bfg\pm\bff ,\ \pm 4\bff\pm 2\bfe\pm\bfg ,\ \pm 4\bff\pm 2\bfg\pm\bfe$,\hfill\hfill&\\
&$\pm 4\bfg\pm 2\bfe\pm\bff ,\ \pm 4\bfg\pm 2\bff\pm\bfe$\hfill\hfill&48\\\hline
$\sqrt{22}$&$\pm 3\bfe\pm 3\bff\pm 2\bfg ,\ \pm 3\bfe\pm 3\bfg\pm 2\bff ,\ \pm 3\bff\pm 3\bfg\pm 2\bfe$\hfill\hfill&24\\\hline
$\sqrt{24}$&$\pm 4\bfe\pm 2\bff\pm 2\bfg ,\ \pm 4\bff\pm 2\bfe\pm 2\bfg ,\ \pm 4\bfg\pm 2\bff\pm 2\bfe$\hfill\hfill&24\\\hline
5&$\pm 5\bfe ,\ \pm 5\bff ,\ \pm 5\bfg ,\ \pm 4\bfe\pm 3\bff ,\ \pm 4\bfe\pm 3\bfg ,\ \pm 4\bff\pm 3\bfe$,\hfill\hfill&\\
&$\pm 4\bff\pm 3\bfg ,\ \pm 4\bfg\pm3\bfe ,\ \pm 4\bfg\pm 3\bff$\hfill\hfill&30\\\hline
$\vdots$&\hfill\hfill&\\\hline
6&$\pm 6\bfe ,\ \pm 6\bff ,\ \pm 6\bfg ,\ \pm 4\bfe\pm 4\bff\pm 2\bfg ,\ \pm 4\bfe\pm 4\bfg\pm 2\bff ,\ \pm 4\bff\pm 4\bfg\pm 2\bfe$\hfill\hfill&30\\
\hline\noalign{\medskip}
\multicolumn{3}{c}{\textbf{Table 1}}\\
\end{tabular}
\parindent=18pt}
\newpage

\begin{tabular}{c|c|c|c|c|c|c|c|c|c|c|c|c|c|c|c|c|c|c}
Total Energy&\multicolumn{2}{|c|}{1}&\multicolumn{5}{|c|}{2}&\multicolumn{9}{|c|}{3}\\
\hline
mass${}^2$&1&0&4&3&2&1&0&9&8&7&6&5&4&3&1&0\\
\hline
number&1&6&1&6&12&8&6&1&6&12&8&6&24&24&12&30\\
\hline
\end{tabular}
\vskip 2pc

\begin{tabular}{c|c|c|c|c|c|c|c|c|c|c|c|c|c|c|c|c|c|c}
Total Energy&\multicolumn{15}{|c|}{4}\\
\hline
mass${}^2$&16&15&14&13&12&11&10&8&7&6&5&4&3&2&0\\
\hline
number&1&6&12&8&6&24&24&12&30&24&24&8&24&48&6\\
\hline
\end{tabular}
\vskip 2pc

{\parindent=-5pc
\begin{tabular}{c|c|c|c|c|c|c|c|c|c|c|c|c|c|c|c|c|c|c|c|c|c|c|c|c|c|c}
Total Energy&\multicolumn{23}{|c|}{5}\\
\hline
mass${}^2$&25&24&23&22&21&20&19&17&16&15&14&13&12&11&9&8&7&6&5&4&3&1&0\\
\hline
number&1&6&12&8&6&24&24&12&30&24&24&8&24&48&6&48&36&24&24&48&24&24&30\\
\hline\noalign{\medskip}
\multicolumn{24}{c}{\textbf{Table 2}}\\
\end{tabular}
\parindent=18pt}
\bigskip

We classify the mass zero particles (photons) into two types, the one dimensional photons (e.g. $\bfe , 2\bff ,\ldots$) that we call \textit{light photons}
(pun intended) and the 2-dimensional and 3-dimensional photons (e.g. $4\bfe +3\bff , 2\bfe +2\bff+\bfg ,\ldots$) that we call \textit{dark photons}. Both of these types have geometric speed~1. However, the light photons propagate along coordinate axes so we can consider their average speed to be 1 while the dark photons do not move directly along coordinate axes so we can consider their average speed to be less than 1. We speculate that these latter particles correspond to dark energy. In a similar way the positive mass particles are either 1, 2 or 3-dimensional and we speculate that the first two types correspond to matter and the last type corresponds to dark matter.

\section{Free Quantum Fields} 
In this section, we study free discrete quantum fields. Since we are considering scalar fields, a particle is essentially determined by its mass. We assume that we are describing a physical system that contains particles of a finite number of various types. For illustrative purposes, suppose there are two types of particles under consideration which we call $p$-particles and $q$-particles with masses $m$ and $M$, respectively, $m\ne M$. The sets
\begin{align*}
\Gamma _m&=\brac{p\in\sscripthat\colon\doubleab{p}_4=m}\\
\Gamma _M&=\brac{q\in\sscripthat\colon\doubleab{q}_4=M}
\end{align*}
are called the mass \textit{hyperboloids} for the particles. To describe these particles quantum mechanically, we construct a complex Hilbert space $K$. Technically speaking,
$K$ is a symmetric Fock space, but the details are not important here. All we need to know is that $K$ exists and has a very descriptive orthonormal basis of the form
\begin{equation*}
\ket{p_1p_2\ldots p_nq_1q_2\ldots q_s}
\end{equation*}
that represents the quantum state in which there are $n$ $p$-particles and $s$ $q$-particles where $p_i,q_j\in\sscripthat$, $i=1,\ldots ,n$, $j=1,\ldots ,s$. The order of the $p$'s and $q$'s is immaterial and different states are mutually orthogonal unit vectors. For example, $\ket{p_1p_2}=\ket{p_2p_1}$ and $\elbows{p_1p_2\mid p_1q_1q_2}=0$. The
\textit{vacuum state} in which there are no particles present is the unit vector $\ket{0}$. The one-particle states of type $p$ are $\ket{p}$, $p\in\Gamma _m$, the two-particle states of the type $p$ are $\ket{p_1p_2}$, etc.

For $p\in\Gamma _m$, we define the \textit{annihilation operator} $a(p)$ on $K$ by $a(p)\ket{0}=0$ where 0 is the zero vector and
\begin{equation*}
a(p)\ket{pp\ldots pp_1\ldots p_nq_1\ldots q_s}=\sqrt{n}\,\ket{pp\ldots pp_1\ldots p_nq_1\ldots q_s}
\end{equation*}
where $n$ $p$'s appear in the vector on the left side and $n-1$ $p$'s appear in the vector on the right side. We interpret $a(p)$ as the operator that annihilates a particle with energy-momentum $p$. For example, $a(p)\ket{p}=\ket{0}$, $a(p)\ket{ppq}=\sqrt{2}\,\ket{pq}$ and
\begin{equation*}
a(q)\ket{p_1p_2qqq}=\sqrt{3}\,\ket{p_1p_2qq}
\end{equation*}
The adjoint $a(p)^*$ of $a(p)$ is the operator on $K$ defined by
\begin{equation*}
a(p)^*\ket{pp\ldots pp_1\ldots p_nq_1\ldots q_s}=\sqrt{n+1}\,\ket{pp\ldots pp_1\ldots p_nq_1\ldots q_s}
\end{equation*}
where $n$ $p$'s appear in the vector on the left side and $n+1$ $p$'s appear in the vector on the right side. For example $a(p)^*\ket{0}=\ket{p}$ and
$a(p)^*\ket{p}=\sqrt{2}\,\ket{pp}$. We interpret $a(p)^*$ as the operator that creates a particle with energy-momentum $p$. To show that $a(p)^*$ is indeed the adjoint of $a(p)$, suppose that
\begin{equation*}
\ket{\alpha}=\ket{pp\ldots pq_1\ldots q_s}
\end{equation*}
has $n$ $p$'s. Then the inner product of $a(p)^*\ket{\alpha}$ with any other basis vector $\ket{\beta}$ is zero unless $\beta$ has the same form as $\ket{\alpha}$ except now there are $n+1$ $p$'s. In this case we have
\begin{equation*}
\bra{\beta}a(p)^*\ket{\alpha}=\sqrt{n+1}\,\elbows{\beta\mid\beta}=\bra{\alpha}a(p)\ket{\beta}
\end{equation*}
which is the defining relationship for the adjoint.

The operators $a(p)$ and $a(p)^*$ have the characteristic property that their commutator $\sqbrac{a(p),a(p)^*}=I$. We illustrate this for two cases which should convince the reader that this holds. If $\ket{p_1\ldots p_nq_1\ldots q_s}$ contains no $p$'s, then
\begin{align*}
a(p)a(p)^*\ket{p_1\ldots p_nq_1\ldots q_s}&=a(p)\ket{pp_1\ldots p_nq_1\ldots q_s}\\
   &=\ket{p_1\ldots p_nq_1\ldots q_s}
\end{align*}
and $a(p)^*a(p)\ket{p_1\ldots p_nq_1\ldots q_s}=0$. Next, if $\ket{\alpha}$ and $\ket{\beta}$ are the vectors in the previous paragraph, then
\begin{equation*}
a(p)a(p)^*\ket{\alpha}=\sqrt{n+1}\,a(p)\ket{\beta}=(n+1)\ket{\alpha}
\end{equation*}
Since $a(p)^*a(p)\ket{\alpha}=n\ket{\alpha}$ we have that $\sqbrac{a(p),a(p)^*}\ket{\alpha}=\ket{\alpha}$.

One reason that $a(p)$ and $a(p)^*$ are important is that they can be employed to describe physically relevant operators. For example the \textit{free total energy operator} for $p$-particles is defined by
\begin{equation*}
P_0=\sum _{p\in\Gamma _m}p_0a(p)^*a(p)
\end{equation*}
The eigenvectors of $P_0$ have the form $\ket{\alpha}=\ket{p_1p_2\ldots p_n}$, $p_j\in\Gamma _m$, $j=1,\ldots ,n$, with eigenvalues the total energy
\begin{equation*}
E_\alpha =(p_1)_0+\cdots +(p_n)_0
\end{equation*}
In a similar way, we define the \textit{free momentum operators}
\begin{equation*}
P_j=\sum _{p\in\Gamma _m}p_ja(p)^*a(p)
\end{equation*}
for $j=1,2,3$. We also define the \textit{number operator}
\begin{equation*}
N=\sum _{p\in\Gamma _m}a(p)^*a(p)
\end{equation*}
Then $N\ket{\alpha}=n\ket{\alpha}$ where $n$ is the number of $p$-particles in the basis vector $\ket{\alpha}$. Of course, all these operators are self-adjoint.

The most important operators in this work are the \textit{free quantum fields}
\begin{equation}         
\label{eq31}
\phi (x)=\sum _{p\in\Gamma _m}\frac{1}{p_0}\sqbrac{a(p)e^{i\pi px/2}+a(p)^*e^{-i\pi px/2}}
\end{equation}
where $x\in\sscript$ and $px$ is the indefinite inner product
\begin{equation*}
px=p_0x_0-p_1x_1-p_2x_2-p_3x_3
\end{equation*}
The self-adjoint operator $\phi (x)$ is an observable representing a quantum field for a particle of mass $m$ at the spacetime point $x$. In standard quantum field theory, the summation in \eqref{eq31} is replaced by an integral over the mass hyperboloid and it is questionable whether such an integral exists. (It is not even clear whether the underlying Fock space $K$ exists.) However, in the discrete case, the summation in \eqref{eq31} does exist. Although this holds in general, we shall illustrate it for a simple, but important case. Denoting the mass hyperboloid for light photons by $\Gamma '_0$ we have that
\begin{equation*}
\phi (0)\ket{0}=\sum _{p\in\Gamma '_0}\frac{1}{p_0}\,\ket{p}
\end{equation*}
Now $\sigma (0)\ket{0}\in K$ with
\begin{equation*}
\doubleab{\phi (0)\ket{0}}^2=6\sum _{n=1}^\infty\frac{1}{n^2}=\pi ^2
\end{equation*}
Hence, $\doubleab{\phi (0)\ket{0}}=\pi$.

Let $\phi (x)$ be a general free quantum field given by \eqref{eq31}. The next result gives an expression for the commutator $\sqbrac{\phi (x),\phi (y)}$.
\begin{thm}       
\label{thm31}
The commutator
\begin{equation*}
\sqbrac{\phi (x),\phi (y)}=2i\sum _{p\in\Gamma _m}\frac{1}{p_0^2}\,\sin\sqbrac{\pi p_0(x_0-y_0)/2}\cos\sqbrac{\bfp\ctimes (\bfx -\bfy )/2}I
\end{equation*}
\end{thm}
\begin{proof}
We have that
\begin{align*}
\phi (x)\phi (y)&=\sum _{p\in\Gamma _m}\frac{1}{p_0}\,\sqbrac{a(p)e^{i\pi px/2}+a(p)^*e^{-i\pi px/2}}\\
  &\quad\ctimes\sum _{q\in\Gamma _m}\frac{1}{q_0}\sqbrac{a(q)e^{i\pi qy/2}+a(q)^*e^{-i\pi qy/2}}\\
  &=\sum _{p,q\in\Gamma _m}\frac{1}{p_0q_0}\left[a(p)a(q)e^{i\pi (px+qy)/2}+a(p)a(q)^*e^{i\pi (px-qy)/2}\right.\\
  &\quad\left. +a(p)^*a(q)e^{i\pi(qy-px)/2}+a(p)^*a(q)^*e^{-i\pi (px+qy)/2}\right]
\end{align*}
Similarly,
\begin{align*}
\phi (y)\phi (x)&=\sum _{p,q\in\Gamma _m}\frac{1}{p_0q_0}\,\left[a(q)a(p)e^{i\pi (px+qy)/2}+a(q)a(p)^*e^{i\pi (qy-px)/2}\right.\\
  &\quad\left. +a(q)a(p)^*e^{i\pi (px-qy)/2}+a(q)^*a(p)^*e^{-i\pi (px+qy)/2}\right]
\end{align*}
Since $\sqbrac{a(p),a(q)}=\sqbrac{a(p)^*,a(q)^*}=0$ and $\sqbrac{a(p),a(q)^*}=\delta _{q,p}I$, we have that
\begin{align*}
\sqbrac{\phi (x),\phi (y)}&=\sum _{p\in\Gamma _m}\frac{1}{p_0^2}\,\brac{\sqbrac{a(p),a(p)^*}e^{i\pi p(x-y)/2}-\sqbrac{a(p),a(p)^*}e^{-i\pi p(x-y)/2}}I\\
  &=2i\sum _{p\in\Gamma _m}\frac{1}{p_0^2}\,\sin\sqbrac{\pi p(x-y)/2}I\\
  &=2i\sum _{p\in\Gamma _m}\frac{1}{p_0^2}\,\sin\sqbrac{\pi p_0(x_0-y_0)/2-\pi\bfp\ctimes (\bfx -\bfy )/2}I\\
  &=2i\sum _{p\in\Gamma _m}\frac{1}{p_0^2}\,\left[\sin\pi p_0(x_0-y_0)/2\cos\pi\bfp\cdot (\bfx -\bfy )/2\right.\\
  &\hskip 6pc\left. -\cos\pi p_0(x_0-y_0)/2\sin\pi\bfp\ctimes (\bfx -\bfy )/2\right]I
\end{align*}
If  $(p_0,\bfp)\in\Gamma _m$, then $(p_0,-\bfp)\in\Gamma _m$ and the result follows because $\sin$ is an odd function.
\end{proof}

The next result is called the \textit{equal-time commutation relation}.
\begin{cor}       
\label{cor32}
If $x_0=y_0$, then $\sqbrac{\phi (x),\phi (y)}=0$.
\end{cor}

\textit{Quantum locality} says that if $\doubleab{x-y}_4^2<0$, then $\sqbrac{\phi (x),\phi (y)}=0$. That is, if $x$ and $y$ are space-like separated, then a field measurement at $x$ cannot affect a field measurement at $y$. If $x_0=y_0$ and $x\ne y$ then $\doubleab{x-y}_4^2<0$ and by Corollary~\ref{cor32}, quantum locality holds. However, in contrast to standard quantum field theory \cite{vel94}, quantum locality does not hold, in general. We show this with the following counterexample. Suppose we have light photons and let $x=(1,2\bfe )$, $y=0$ so that $\doubleab{x-y}_4^2=-3<0$. Applying Theorem~\ref{thm31} we have that
\begin{align*}
\sqbrac{\phi (x),\phi (y)}&=2i\sum _{p\in\Gamma '_0}\frac{1}{p_0^2}\,\sin (\pi p_0/2)\cos\pi\bfp\ctimes\bfe I\\
  &=4i\sum _{n=1}^\infty\frac{1}{n^2}\,\sin (\pi n/2)\cos\pi nI=4i\sum _{n=1}^\infty\frac{(-1)^n}{n^2}\,\sin\pi n/2I\\
  &=-4i\sum _{n=1}^\infty\,\frac{1}{(2n-1)^2}\,\sin (2n-1)\pi /2I=4i\sum _{n=1}^\infty\frac{(-1)^n}{(2n-1)^2}\,I\ne 0
\end{align*}

\section{Interacting Quantum Fields} 
The important part of quantum fields occurs when we have interactions because then we obtain nontrivial scattering. This section considers interaction Hamiltonians and scattering operators. These are relevant because most of modern theoretical and experimental physics involves some kind of scattering. We shall eventually construct some examples of interaction Hamiltonians, but for now let $H(x_0)$ be self-adjoint operators on $K$ that describe an interaction, where $x_0=0,1,2,\ldots$, represents time. The corresponding \textit{scattering operators} $S(x_0)$ satisfy a ``second quantization'' equation
\begin{equation}         
\label{eq41}
\nabla _{x_0}S(x_0)=iH(x_0)S(x_0)
\end{equation}
Of course, \eqref{eq41} is a generalization of Schr\"odinger's equation and for this work $\nabla _{x_0}$ is the difference operator
\begin{equation*}
\nabla _{x_0}S(x_0)=S(x_0+1)-S(x_0)
\end{equation*}
Starting with the initial condition $S(0)=I$, we obtain from \eqref{eq41} that
\begin{equation*}
S(1)=I+iH(0)
\end{equation*}
Continuing, we conclude that
\begin{align}        
\label{eq42}
S(2)&=S(1)+iH(1)S(1)=\sqbrac{I+iH(1)}S(1)=\sqbrac{I+iH(1)}\sqbrac{I+iH(0)}\notag\\
S(3)&=S(2)+iH(2)S(2)=\sqbrac{I+iH(2)}S(2)=\sqbrac{I+iH(2)}\sqbrac{I+iH(1)}\sqbrac{I+iH(0)}\notag\\
\vdots&\notag\\
S(n)&=\sqbrac{I+iH(n-1)}\sqbrac{I+iH(n-2)}\cdots\sqbrac{I+iH(1)}\sqbrac{I+iH(0)}
\end{align}

In general, the operators $H(j)$ do not commute so the order in \eqref{eq42} must be retained. This is called a \textit{time ordered product} of operators. Most of the complication in quantum field theory results from trying to solve \eqref{eq41}. In fact, it appears that solving \eqref{eq41} exactly is intractable in general and all we can accomplish is to solve \eqref{eq42} approximately using ``perturbation'' techniques. For $n\ne 0$, $S(n)$ is not unitary, in general. However, presumably the \textit{limiting scattering operator}
$S=\lim\limits _{n\infty}S(n)$ should be unitary in order to preserve probability. Of course, this depends on the interaction Hamiltonian $H(n)$. The author does not know reasonable conditions on $H(n)$ that ensure the unitarity of $S$.

If we multiply \eqref{eq42} out, we obtain the useful form
\begin{align}        
\label{eq43}
S(n)&=I+i\sum _{j=0}^{n-1}H(j)+i^2\sum _{j_2<j_1}^{n-1}H(j_1)H(j_2)+i^3\sum _{j_3<j_2<j_1}^{n-1}H(j_1)H(j_2)H(j_3)\notag\\
&\qquad +\cdots +i_nH(n-1)H(n-2)\cdots H(0)
\end{align}
Equation~\eqref{eq43} gives a quantum inclusion-exclusion principle with the interaction Hamiltonian again time ordered. The interaction Hamiltonian $H(x_0)$ is usually constructed from an \textit{interaction Hamiltonian density} consisting of self-adjoint operators on $K$ denoted by $\hscript (x)$, $x\in\sscript$. Letting
\begin{equation*}
V(x_0)=\ab{\brac{\bfx\in\integers ^3\colon\doubleab{\bfx}_3\le x_0}}
\end{equation*}
be the cardinality of the set in brackets (called the \textit{space volume} at $x_0$) we define
\begin{equation}        
\label{eq44}
H(x_0)=\frac{1}{V(x_0)}\sum\brac{\hscript (x_0,\bfx )\colon\doubleab{\bfx}_3\le x_0}
\end{equation}
The first few terms of \eqref{eq44} are: $H(0)=\hscript (0)$
\begin{align*}
H(1)&=\tfrac{1}{7}\sqbrac{\hscript (1,\bfzero )+\hscript (1,\bfe )+\hscript (1,-\bfe )+\hscript (1,\bff )+\hscript (1,-\bff )+H(1,\bfg )+H(1,-\bfg )}\\
H(2)&=\tfrac{1}{33}\sqbrac{\hscript (2,\bfzero )+\hscript (2,\bfe )+\cdots +\hscript (2,2\bfg )+\hscript (2,-2\bfg )}
\end{align*}

We now illustrate this theory with a simplified example that contains the essential elements that can be generalized to more complicated and realistic situations. We consider the scattering of two electrons with mass $m$ and initial energy-momentum $p,q\in\sscripthat$, $p\ne q$. We assume that the electrons interact by exchanging light photons and arrive with final energy-momentum $p',q'\in\sscripthat$. As usual in this article we are neglecting spin so we are really describing spin-zero particles like neutral pions. We say that the \textit{input state} is $\ket{p,q}\in K$ and the \textit{output state} is $\ket{p'q'}\in K$. Let $\phi (x)$, $\sigma (x)$ be the quantum fields for the electrons and light photons, respectively
\begin{align*}
\phi (x)&=\sum _{p\in\Gamma _m}\frac{1}{p_0}\sqbrac{a(p)e^{i\pi px/2}+a(p)^*e^{-i\pi px/2}}\\
\sigma (x)&=\sum _{k\in\Gamma '_0}\frac{1}{k_0}\sqbrac{a(k)e^{i\pi kx/2}+a(k)^*e^{-i\pi kx/2}}
\end{align*}

The interaction Hamiltonian density $\hscript (x)$ is frequently constructed by interacting the fields $\phi$ and $\sigma$. A linear combination of $\phi$ and $\sigma$ will not produce any scattering so we take a simple nonlinear combination of the form \cite{vel94}
\begin{equation}        
\label{eq45}
\hscript (x)=g\phi (x)^2\sigma (x)
\end{equation}
where $g$ is called the \textit{coupling constant}. Ideally, we can now find the scattering operator $S$ and compute the \textit{scattering amplitude} $\bra{p'q'}S\ket{pq}$. The probability of the interaction becomes
\begin{equation*}
P\paren{\ket{pq}\to\ket{p'q'}}=\ab{\bra{p'q'}S\ket{pq}}^2
\end{equation*}
Unfortunately, we cannot find $S$ exactly so we must be content with computing the lower level perturbation terms, $\bra{p'q'}S(n)\ket{pq}$, $n=0,1,2,\ldots\,$. (We shall stop at $n=2$. The furthest anyone usually can go is $n=5$.) As we shall see, these terms contain a considerable amount of symmetry which could imply conservation of energy-momentum. However, we have not proved this so we shall postulate the conservation law
\begin{equation}        
\label{eq46}
p'+q'=p+q
\end{equation}
As one would expect, \eqref{eq46} gives some simplifications.

Applying \eqref{eq43} with $n=2$ we have that
\begin{align}        
\label{eq47}
\bra{p'q'}S(2)\ket{pq}&=\elbows{p'q'\mid pq}+i\bra{p'q'}H(0)\ket{pq}\notag\\
  &\qquad +i\bra{p'q'}H(1)\ket{pq}-\bra{p'q'}H(1)H(0)\ket{pq}
\end{align}

As is usually done, we shall assume that $\ket{p'q'}\ne\ket{pq}$ \cite{ps95,vel94}. We have three reasons for this assumption. One is that $\ket{pq}\to\ket{pq}$ is not of interest because there is no scattering in this case. Another is that this case is unlikely so it would have small probability. Finally, as we shall later show, this case is very hard to compute. Because of this assumption, the first term in \eqref{eq47} vanishes. The second and third terms contain one $H(n)$ and hence only one $\sigma$-field. This applied to a state
$\ket{pq}$ that does not have a $\sigma$-particle (light photon) gives 0 for the $a(k)$ part or a state containing a $\sigma$-particle for the $a(k)^*$ part. The inner product of such a state with the state $\ket{p'q'}$ containing no $\sigma$-particle is 0. Similarly, any product of an odd number of $H(n)$ gives zero between states without $\sigma$-particles.

We conclude that \eqref{eq47} reduces to
\begin{equation}        
\label{eq48}
\bra{p'q'}S(2)\ket{pq}=-\bra{p'q'}H(1)H(0)\ket{pq}
\end{equation}
If we continue to the next order of perturbation, we obtain
 \begin{align}        
\label{eq49}
\bra{p'q'}S(3)\ket{pq}&=-\bra{p'q'}H(1)H(0)\ket{pq}-\bra{p'q'}H(2)H(0)\ket{p}\notag\\
 &\qquad -\bra{p'q'}H(2)H(1)\ket{pq}
\end{align}
As shown earlier, the next term in \eqref{eq43} does not appear in \eqref{eq49}. With some more work we could compute \eqref{eq49} instead of \eqref{eq48}, but to save space we shall only consider \eqref{eq48}. The operator in \eqref{eq48} has the form
\begin{align}        
\label{eq410}
H(1)H(0)&=\frac{g^2}{7}\sum _{x,y}\phi (x)^2\sigma (x)\phi (y)^2\sigma (y)\notag\\
  &=\frac{g^2}{7}\sum _{x,y}\phi (x)^2\phi (y)^2\sigma (x)\sigma (y)
\end{align}
In order to get a nonzero inner product in \eqref{eq48}, we must have that $H(1)H(0)\ket{pq}=\alpha\ket{pq}$ for some $\alpha\in\complex$, $\alpha\ne 0$, which we write
$\ket{pq}\to\ket{p'q'}$. As far as $\sigma (x)\sigma (y)$ is concerned, we only have the possibility $a(k)a(k)^*$, $k\in\Gamma '_0$ which corresponds to an exchange of photons. For $\phi (x)^2\phi (y)^2$ we have six possibilities. These can be described by six Feynman diagrams, but we shall employ a symbolic notation.

We could first annihilate $p$ and $q$ using $\phi (y)^2$ and then create $p'$ and $q'$ using $\phi (x)^2$. Another possibility is to first annihilate $p$ and create $p'$ using
$\phi (y)^2$ and then annihilate $q$ and create $q'$ using $\phi (x)^2$. The six cases can be described symbolically as follows, with the first two cases given above.
{\obeylines
(Case 1)\enspace$\ket{pq}\to\ket{0}\to\ket{p'q'}$
(Case 2)\enspace$\ket{pq}\to\ket{p'q}\to\ket{p'q'}$
(Case 3)\enspace$\ket{pq}\to\ket{q'q}\to\ket{p'q'}$
(Case 4)\enspace$\ket{pq}\to\ket{pp'}\to\ket{p'q'}$
(Case 5)\enspace$\ket{pq}\to\ket{pq'}\to\ket{p'q'}$
(Case 6)\enspace$\ket{pq}\to\ket{pqp'q'}\to\ket{p'q'}$
}\bigskip

Note that although we have conservation of energy-momentum \eqref{eq46} for the input and output states of $\ket{pq}\to\ket{p'q'}$, we do not conserve energy-momentum at intermediate times in the interaction. For example, in Case~1 we do not have $p+q=0$. During the interaction, energy-momentum is annihilated and created by the quantum fields.

We now explain specifically why we assume that $\ket{p'q'}\ne\ket{pq}$. In the situation in which the input and output states are both $\ket{pq}$ we have cases like
\begin{align*}
\ket{pq}&\to\ket{p'q}\to\ket{pq}\\
\ket{pq}&\to\ket{pq'}\to\ket{pq}\\
\ket{pq}&\to\ket{pqp'q'}\to\ket{pq}
\end{align*}
But now $p',q'$ are arbitrary elements of $\Gamma _m$ so we would have to sum over these infinite number of elements which is quite difficult.

The next lemma shows that for a fixed input state $\ket{pq}$, there are only a finite number of output states $\ket{p'q'}$ satisfying $\ket{pq}\to\ket{p'q'}$. In this way we can sum probabilities to get various alternatives.

\begin{lem}       
\label{lem41}
There exist only finitely many $\ket{p'q'}$ such that $\ket{pq}\to\ket{p'q'}$.
\end{lem}
\begin{proof}
If $\ket{pq}\to\ket{p'q'}$, then by \eqref{eq46} $p'+q'=p+q$. Hence,
\begin{equation*}
p'_0,q'_0\le p'_0+q'_0=p_0+q_0
\end{equation*}
We conclude that there are only a finite number of possible values for $p'_0,q'_0$. Since for $p'\in\Gamma _m$ we have that $\doubleab{\bfp '}_3^2=(p'_0)^2-m^2$, there are only a finite number of possible $\bfp '$ and hence only a finite number of possible $p'$.
\end{proof}

Notice that Lemma~\ref{lem41} is general and does not depend on a particular mass $m$. If $\ket{p'q'}\ne\ket{pq}$, then we have seen that there are only six possible cases for each of the finite number of scattering situations. We thus have only a finite number of terms to compute.

\section{Scattering Examples} 
For our first example, we consider the simplified electron scattering process discussed in Section~4. The interaction Hamiltonian density is given by \eqref{eq45} and we want to compute the first order perturbation scattering amplitude \eqref{eq48}. Writing \eqref{eq410} in detail, we have that
\begin{align*}
H(1)H(0)&=\frac{1}{7}\,\sqbrac{\hscript (1,\bfzero )+\hscript (1,\bfe )+\cdots +\hscript (1,-\bfg )}\hscript (0)\\
  &=\frac{g^2}{7}\left[\phi (1,\bfzero )^2\sigma (1,\bfzero )+\phi (1,\bfe )^2\sigma (1,\bfe )\right.\\
  &\hskip 3pc\left. +\cdots +\phi (1,-\bfg )^2\sigma (1,-\bfg )\right]\phi (0)^2\sigma (0)\\
  &=\frac{g^2}{7}\left[\phi (1,\bfzero )^2\phi (0)^2\sigma (1,\bfzero )\sigma (0)+\phi (1,\bfe )^2\phi (0)^2\sigma (1,\bfe )\sigma (0)\right.\\
  &\hskip 3pc\left. +\cdots +\phi (1,-\bfg )^2\phi (0)^2\sigma (1,-\bfg)\sigma (0)\right]
\end{align*}
Taking the photon field $\sigma$ first, we have that
\begin{align*}
\sigma (1,\bfzero )\sigma (0)&=\sum _{k\in\Gamma '_0}\frac{1}{k_0}\,\sqbrac{a(k)e^{i\pi k_0/2}+a(k)^*e^{-i\pi k_0/2}}\\
  &\hskip 3pc\ctimes\sum _{k'\in\Gamma '_0}\frac{1}{k'_0}\sqbrac{a(k')+a(k')^*}
\end{align*}
By our argument in Section~4, $\sigma (1,\bfzero )\sigma (0)$ contributes the coefficient
\begin{equation*}
c_1=6\sum _{n=1}^\infty\frac{1}{n^2}\,e^{i\pi n/2}
\end{equation*}
The next term is
\begin{align*}
\sigma (1,\bfe )\sigma (0)&=\sum _{k\in\Gamma '_0}\frac{1}{k_0}\,\sqbrac{a(k)e^{i\pi (k_0-k_1)/2}+a(k)^*e^{-i\pi (k_0-k_1)/2}}\\
  &\hskip3pc\ctimes\sum _{k'\in\Gamma '_0}\frac{1}{k'_0}\,\sqbrac{a(k')+a(k')^*}
\end{align*}
Since $k_1=\pm k_0$, this term contributes the coefficient
\begin{align*}
c_2&=\sum _{k\in\Gamma '_0}\frac{1}{k_0^2}\,e^{i\pi (k_0-k_1)/2}=\sum _{n=1}^\infty\frac{1}{n_2}\,(1+e^{i\pi n}+4e^{i\pi n/2})\\
  &=\frac{\pi ^2}{12}\,+4\sum _{n=1}^\infty\frac{1}{n^2}\,e^{i\pi n/2}
\end{align*}
Since the other terms contribute this same coefficient, the nonzero part of $H(1)H(0)$ in \eqref{eq48} becomes
\begin{align}        
\label{eq51}
H(1,0)&=\frac{g^2}{7}\left[c_1\phi (1,\bfzero )^2\phi (0)^2+c_2\phi (1,\bfe )^2\phi (0)^2\right.\notag\\
&\hskip 3pc\left. +\cdots +c_2\phi (1,-\bfg )^2\phi (0)^2\right]
\end{align}

We now consider the terms of $H(1,0)$ in \eqref{eq51}. We have that
\begin{align}        
\label{eq52}
\phi (1,\bfzero )^2\phi (0)&=\brac{\sum _{p\in\Gamma _m}\frac{1}{p_0}\,\sqbrac{a(p)e^{i\pi p_0/2}+a(p)^*e^{-i\pi p_0/2}}}^2\notag\\
  &\hskip 3pc\ctimes\brac{\sum _{p\in\Gamma _m}\frac{1}{p_0}\,\sqbrac{a(p)+a(p)^*}}
\end{align}
For the six cases, this term contributes the following coefficients where\newline $\beta = 1/p_0q_0p'_0q'_0$.\bigskip

{\obeylines
(Case 1)\enspace$\beta e^{-i\pi (p'_0+q'_0)/2}$
(Case 2)\enspace$\beta e^{i\pi (q_0-q'_0)/2}$
(Case 3)\enspace$\beta e^{i\pi (q_0-p'_0)/2}$
(Case 4)\enspace$\beta e^{i\pi (p_0-q'_0)/2}$
(Case 5)\enspace$\beta e^{i\pi (p_0-p'_0)/2}$
(Case 6)\enspace$\beta e^{i\pi (p_0+q_0)/2}$
}\bigskip

\noindent Adding these six coefficients and applying conservation of energy-momentum gives the following contribution of \eqref{eq52}
\begin{equation*}
d_1=2\beta\sqbrac{\cos\frac{\pi}{2}\,(p_0+q_0)+\cos\frac{\pi}{2}\,(q_0-q'_0)+\cos\frac{\pi}{2}\,(q_0-p'_0)}
\end{equation*}

The next term in \eqref{eq51} is
\begin{align*} 
\phi (1,\bfe )^2\phi (0)^2&=\brac{\sum _{p\in\Gamma _m}\frac{1}{p_0}\,\sqbrac{a(p)e^{i\pi (p_0-p_1)/2}+a(p)^*e^{-i\pi (p_0-p_1)/2}}}^2\notag\\
  &\hskip 3pc\ctimes\brac{\sum _{p\in\Gamma _m}\frac{1}{p_0}\,\sqbrac{a(p)+a(p)^*}}^2
\end{align*}
Again, we have the six cases:\bigskip

{\obeylines
(Case 1)\enspace$\beta e^{-i\pi (p'_0+q'_0-p'_1-q'_1)/2}$
(Case 2)\enspace$\beta e^{i\pi (q_0-q'_0-q_1+q'_1)/2}$
(Case 3)\enspace$\beta e^{i\pi (q_0-p'_0-q_1+p'_1)/2}$
(Case 4)\enspace$\beta e^{i\pi (p_0-q'_0-p_1+q'_1)/2}$
(Case 5)\enspace$\beta e^{i\pi (p_0-p'_0-p_1+p'_1)/2}$
(Case 6)\enspace$\beta e^{i\pi (p_0+q_0-p_1-q_1)/2}$
}\bigskip

\noindent The sum of these coefficients gives:
\begin{align*}
d_2&=2\beta\left[\cos\frac{\pi}{2}\,(p_0+q_0-p_1-q_1)+\cos\frac{\pi}{2}\,(q_0-q'_0-q_1+q'_1)\right.\\
  &\hskip 3pc\left. +\cos\frac{\pi}{2}\,(q_0-p'_0-q_1+p'_1)\right]
\end{align*}

The next term in \eqref{eq51} has the form $\phi (1,-\bfe )^2\phi (0)^2$ and the sum of the corresponding coefficients $d_3$ will be the same as $d_2$ except the $p_1,q_1,p'_1,q'_1$ terms will be the negatives of those in $d_3$. We then have
\begin{align*}
d_2+d_3&=4\beta\left[\cos\frac{\pi}{2}\,(p_0+q_0)\cos\frac{\pi}{2}\,(p_1+q_1)\right.\\
  &\left. +\cos\frac{\pi}{2}\,(q_0-q'_0)\cos\frac{\pi}{2}\,(-q_1+q'_1)+\cos\frac{\pi}{2}\,(q_0-p'_0)\cos\frac{\pi}{2}\,(-q_1+p'_1)\right]\end{align*}
The other terms will be similar, so adding all these terms gives
\begin{align*}
&\bra{p'q'}S(2)\ket{pq}=\frac{-g^2}{7}\,\sqbrac{c_1d_1+c_2(d_2+d_3+d_4+d_5+d_6)}\\
&=\frac{-2g^2}{7}\,\beta\left\{\cos\frac{\pi}{2}\,(p_0+q_0)\sqbrac{c_1+2c_2\paren{\cos\frac{\pi}{2}\,(p_1+q_1)+\cos\frac{\pi}{2}\,(p_2+q_2)+\cos\frac{\pi}{2}\,(p_3+q_3)}}\right.\\
\noalign{\medskip}
&\quad +\cos\frac{\pi}{2}\,(q_0-q'_0)\sqbrac{c_1+2c_2\paren{\cos\frac{\pi}{2}\,(q'_1-q_1)+\cos\frac{\pi}{2}\,(q'_2-q_2)+\cos\frac{\pi}{2}\,(q'_3-q_3)}}\\
\noalign{\medskip}
&\left.\quad +\cos\frac{\pi}{2}\,(q_0-p'_0)\sqbrac{c_1+2c_2\paren{\cos\frac{\pi}{2}\,(p'_1-q_1)+\cos\frac{\pi}{2}\,(p'_2-q_2)+\cos\frac{\pi}{2}\,(p'_3-q_3)}}\right\}\\
\end{align*}

We now consider a specific example. Let $p=(2,1,1,1)$, $q=(3,-2,-2,0)$, $p'=(2,-1,1,-1)$ and $q'=(3,0,-2,2)$. In this case, the mass $m=1$ and we have conservation of energy-momentum because
\begin{align*}
p+q&=p'+q'=(5,-1,-1,1)\\
\intertext{Moreover,}
\beta&=\frac{1}{p_0q_0p'_0q'_0}=\frac{1}{36}
\end{align*}
Applying our previous formula, the amplitude becomes
\begin{equation*}
\bra{p'q'}S(2)\ket{pq}=\frac{-g^2}{126}\,\sqbrac{c_1+2c_2(-1+1-1)}=\frac{g^2}{126}\,(2c_2-c_1)
\end{equation*}
To the first order perturbation we have that
\begin{equation*}
P\paren{\ket{pq}\to\ket{p'q'}}=\ab{\bra{p'q'}S(2)\ket{pq}}^2=\frac{g^4}{(126)^2}\,(2c_2-c_1)^2
\end{equation*}
Since $g$ is unknown, this does not tell us anything. However, we can use this calculation to compare probabilities. Let $p,q$ be as before and let $p''=(2,1,-1,-1)$,
$q''=(3,-2,0,2)$. We again obtain
\begin{equation*}
\bra{p''q''}S(2)\ket{pq}=\frac{g^2}{126}\,(2c_2-c_1)
\end{equation*}
which is not surprising due to the symmetry of the situation. It does however, exhibit some consistency. To illustrate a more interesting comparison, suppose the initial state is
$p=(3,2,2,0)$, $q=(3,-2,-2,0)$ and the final state is $p'=(3,0,2,2)$, $q'=(3,0,-2,-2)$. We then obtain
\begin{equation*}
\bra{p'q'}S(2)\ket{pq}=\frac{g^2}{126}\,(8c_2-c_1)
\end{equation*}
which is quite different from what we obtained before.

Our second example is a simplified version of electron-proton scattering due to the exchange of photons. As before, we have the electron and photon fields $\phi (x)$,
$\sigma (x)$, but now we include a proton field of mass $M$
\begin{equation*}
\psi (x)=\sum _{p\in\Gamma _M}\sqbrac{a(p)e^{i\pi px/2}+a(p)^*e^{-i\pi px/2}}
\end{equation*}
Let the interaction Hamiltonian density be $\hscript =g\phi ^2\sigma -g\psi ^2\sigma$ where the negative sign is because the electron and proton have opposite charge. Let
$p\in\Gamma _m$ be an initial electron, $q\in\Gamma _M$ be as initial proton and we consider the scattering process $\ket{pq}\to\ket{p'q'}$ where
$\ket{p'q'}\ne\ket{pq}$ is the final state. As before, we study the first perturbation term \eqref{eq48} where $H(1)H(0)$ now has the form
\begin{align*}
H(1)H(0)&=\frac{g^2}{7}\,\sum _{x,y}\sqbrac{\phi (x)^2\sigma (x)-\psi (x)^2\sigma (x)}\sqbrac{\phi (y)^2\sigma (y)-\psi (y)^2\sigma (y)}\\
  &=\frac{g^2}{7}\,\sum _{x,y}\sqbrac{\phi (x)^2\phi (y)^2-\phi (x)^2\psi (y)^2-\psi (x)^2\phi (y)^2+\psi (x)^2\psi (y)^2}\sigma (x)\sigma (x)
\end{align*}
To map $\ket{pq}$ to $\ket{p'q'}$, we again have the term $a(k)a(k)^*$ for $k\in\Gamma '_0$ and we obtain the same two constants $c_1$ and $c_2$. For the electron-proton terms we have only two cases:\bigskip

{\obeylines
(Case 1')\enspace$\ket{pq}\to\ket{pq'}\to\ket{p'q'}$
(Case 2')\enspace$\ket{pq}\to\ket{p'q}\to\ket{p'q'}$
}\bigskip

As in \eqref{eq51}, the nonzero part of $H(1)H(0)$ is
\begin{align*}
H(1,0)&=\frac{-g^2}{7}\,\left[c_1\phi (1,\bfzero )^2\psi (0)^2+c_1\psi (1,\bfzero )^2\phi (0)^2+c_2\phi (1,\bfe )^2\psi (0)^2\right.\\
  &\quad\left. +c_2\psi (1,\bfe )^2\phi (0)^2+\cdots +c_2\phi (1,-\bfg )^2\psi (0)^2+c_2\psi (1,-\bfg )^2\phi (0)^2\right]
\end{align*}
The first term on the right side gives
\begin{align*}
\phi (1,\bfzero )^2\psi (0)^2&=\brac{\sum _{p\in\Gamma _m}\frac{1}{p_0}\,\sqbrac{a(p)e^{i\pi p_0/2}+a(p)^*e^{-i\pi p_0/2}}}^2\\
  &\qquad\ctimes\brac{\sum _{p\in\Gamma _M}\frac{1}{q_0}\,\sqbrac{a(q)+a(q)^*}}^2
\end{align*}
This applies to Case~1' and contributes the coefficient
\begin{equation*}
\beta e^{i\pi (p_0-p'_0)/2}
\end{equation*}
In a similar way, the second term on the right side applies to Case~2' and contributes the coefficient
\begin{equation*}
\beta e^{i\pi (q_0-q'_0)/2}
\end{equation*}
The third term contributes
\begin{equation*}
\beta e^{i\pi (p_0-p'_0)/2}e^{i\pi (p'_1-p_1)/2}
\end{equation*}
while the fourth term contributes
\begin{equation*}
\beta e^{i\pi(q_0-q'_0)/2}e^{i\pi (q'_1-q_1)/2}
\end{equation*}
The other terms are similar. Adding all these terms gives
\begin{align*}
\bra{p'q'}S(2)\ket{pq}&=\frac{-g^2\beta}{7}\,\left\{e^{i\pi (p_0-p'_0)/2}\sqbrac{c_1+2c_2\sum _{j=1}^3\cos\frac{\pi}{2}\,(p_j-p'_j)}\right.\\
&\qquad\left. +e^{i\pi (q_0-q'_0)/2}\sqbrac{c_1+2c_2\sum _{j=1}^3\cos\frac{\pi}{2}(q_j-q'_j)}\right\}
\end{align*}
Applying conservation of energy-momentum, this becomes
\begin{equation}        
\label{eq53}
\bra{p'q'}S(2)\ket{pq}=\frac{-2g^2\beta}{7}\,\brac{\cos\frac{\pi}{2}\,(p_0-p'_0)\sqbrac{c_1+2c_2\sum _{j=1}^3\cos\frac{\pi}{2}\,(p_j-p'_j)}}
\end{equation}\smallskip

\noindent Notice that \eqref{eq53} is independent of $q$ and $q'$. This may not be true for higher order perturbations such as $S(3)$.

By Lemma~\ref{lem41}, we know that there are only a finite number of possible final states $\ket{p^jq^j}$ for the initial state $\ket{pq}$. As usual, we assume that
$\rmprob\paren{\ket{pq}\to\ket{pq}}$ is small. The \textit{flux} is the number of particles, per unit surface area, per unit time and this is given by the geometric speed
\begin{equation*}
\frac{\doubleab{\bfp ^j}_3}{p_0^j}+\frac{\doubleab{\bfq ^j}_3}{q_0^j}
\end{equation*}
The \textit{cross section} $\sigma$ is defined to be the probability, per unit time, for unit flux summed over the final states. Since the total time, in our case, is 2, we have that
\begin{align}        
\label{eq54}
\sigma&=\frac{1}{2}\,\sum _j\sqbrac{\frac{\doubleab{\bfp ^j}_3}{p_0^j}+\frac{\doubleab{\bfq ^j}_3}{q_0^j}}\rmprob\paren{\ket{pq}\to\ket{p^jq^j}}\notag\\
&=\frac{1}{2}\,\sum _j\sqbrac{\frac{\doubleab{\bfp ^j}_3}{p_0^j}+\frac{\doubleab{\bfq ^j}_3}{q_0^j}}\ab{\bra{p^jq^j}S(2)\ket{pq}}^2
\end{align}

We now restrict our attention to cases of the form $p=(p_0,0,0,p_3)$, $q=(q_0,0,0,0)$. Thus, electrons are moving along the $z$-axis and the proton is initially stationary. Furthermore, we make the approximation that $m=0$ so that $p_3=p_0$ and the electron acts like a photon. This a good approximation because the electron is much less massive than the proton (about 1 to 2000). To be specific, consider the simple example $p=(5,0,0,5)$, $q=(5,0,0,0)$. It is easy to check that conservation of energy-momentum implies that the only possible outgoing state $\ket{p'q'}\ne\ket{pq}$ is given by $p'=(3,2,2,1)$, $q'=(7,-2,-2,4)$. Substituting these values into \eqref{eq53} gives
\begin{equation*}
\bra{p'q'}S(2)\ket{pq}=\frac{2g^2\beta}{7}\,(c_1+2c_2)
\end{equation*}
Applying \eqref{eq54}, the cross section becomes
\begin{equation*}
\sigma =\frac{2}{49}\,g^4\beta ^2(c_1+2c_2)^2
\end{equation*}

Finally, we briefly consider particle decay rates and lifetimes. The decay probability per unit time is the \textit{decay rate} and the inverse of the decay rate is the
\textit{lifetime}. Suppose we have $\phi$-particles and $\sigma$-particles, but now we assume that the $\sigma$-particle is more massive than two $\phi$-particles so the
$\sigma$ decays into two $\phi$'s. We again consider an interaction Hamiltonian density $\hscript =g\phi ^2\sigma$. Let $k$ the initial energy-momentum of the $\sigma$ and
$p,q$ the final energy-momentum of the two $\phi$-particles. At the lower order perturbation, we have only one case for the decay, namely $\ket{k}\to\ket{pq}$. The last term of the scattering operator
\begin{equation*}
S(2)=I+i\sqbrac{H(0)+H(1)}-H(1)H(0)
\end{equation*}
gives $\bra{pq}H(1)H(0)\ket{k}=0$ and we have
\begin{align}        
\label{eq55}
\bra{pq}S(2)\ket{k}&=i\sqbrac{\bra{pq}H(0)\ket{k}+\bra{pq}H(1)\ket{k}}\notag\\
  &=ig\biggl\{\bra{pq}\phi (0)^2\sigma (0)\ket{k}+\frac{1}{7}\left[\bra{pq}\phi (1,\bfzero )^2\sigma (1,\bfzero )\ket{k}\right.\biggr.\notag\\
  &\quad\biggl.\left. +\bra{pq}\phi (1,\bfe )^2\sigma (1,\bfe )\ket{k}+\cdots +\bra{pq}\phi (1,-\bfg )^2\sigma (1,-\bfg )\ket{k}\right]\biggr\}\notag\\
  &=\frac{ig}{k_0p_0q_0}\,\biggl\{1+\frac{1}{7}\,e^{-i\pi (p_0+q_0)/2}\left[ e^{i\pi k_0/2}+e^{i\pi (k_0-k_1)/2}e^{i\pi (p_1+q_1)/2}\right.\biggr.\notag\\
  &\hskip 4pc\biggl.\left. +\cdots +e^{i\pi (k_0-k_3)/2}e^{i\pi (p_3+q_3)/2}\right]\biggr\}
\end{align}

Now suppose we are in the $\sigma$ rest system so that $k=k_0$ and $\bfk =0$. By conservation of energy-momentum, $p_0+q_0=k_0$ and $\bfp +\bfq =\bfzero$. Then
$\bfp =-\bfq$ so $q_0=p_0$ and $k_0=2p_0$. Hence, \eqref{eq55} simplifies to
\begin{equation}        
\label{eq56}
\bra{pq}S(2)\ket{k}=\frac{8ig}{k_0^3}
\end{equation}
We can now use \eqref{eq56} for the first approximation to decay rates and lifetimes.


\begin{thebibliography}{99}
\bibitem{gud68}S.~Gudder, Elementary length topologies in physics, \textit{SIAM J.\ Appl.\ Math.} \textbf{16}, 1011--1019 (1968).
\bibitem{gud16}S.~Gudder, Discrete quantum gravity and quantum field theory, arXiv: gr-qc 1603.03471v1 (2016).
\bibitem{hei30}W.~Heisenberg, \textit{The Physical Principles of Quantum Mechanics}, University of Chicago Press, Chicago (1930).
\bibitem{ps95}M.~Peskin and D.~Schroeder, \textit{An Introduction to Quantum Field Theory}, Addison-Wesely, Reading, Mass. (1995).
\bibitem{rus54}B.~Russell, \textit{The Analysis of Matter}, Dover, New York (1954).
\bibitem{sw64}R.~Streater and A.~Wightmann, \textit{PCT, Spin and Statistics and all that}, Benjamin, New York (1964).
\bibitem{vde87}C.~Vanden Eynden, \textit{Elementary Number Theory}, McGraw Hill, Boston, Mass. (1987).
\bibitem{vel94}M.~Veltman, \textit{Diagrammatica}, Cambridge University Press, Cambridge (1994).
\end{thebibliography}
\end{document}